**Geographic Space as a Living Structure for Predicting Human Activities Using Big Data**

Bin Jiang and Zheng Ren

Faculty of Engineering and Sustainable Development, Division of GIScience
University of Gävle, SE-801 76 Gävle, Sweden
Email: bin.jiang@hig.se, renzheng1991@gmail.com

*I propose a view of physical reality which is dominated by the existence of this one particular structure, W, the wholeness. In any given region of space, some subregions have higher intensity as centers, others have less. Many subregions have weak intensity or none at all. The overall configuration of the nested centers, together with their relative intensities, comprise a single structure. I define this structure as "the" wholeness of that region.*

Christopher Alexander (2002-2005, Book 1, P. 96)

**Abstract**
Inspired by Christopher Alexander's conception of the world – space is not lifeless or neutral, but a living structure involving far more small things than large ones – a topological representation has been previously developed to characterize the living structure or the wholeness of geographic space. This paper further develops the topological representation and living structure for predicting human activities in geographic space. Based on millions of street nodes of the United Kingdom extracted from OpenStreetMap, we established living structures at different levels of scale in a nested manner. We found that tweet locations at different levels of scale, such as country and city, can be well predicted by the underlying living structure. The high predictability demonstrates that the living structure and the topological representation are efficient and effective for better understanding geographic forms. Based on this major finding, we argue that the topological representation is a truly multiscale representation, and point out that existing geographic representations are essentially single scale, so they bear many scale problems such as modifiable areal unit problem, the conundrum of length，and the ecological fallacy. We further discuss on why the living structure is an efficient and effective instrument for structuring geospatial big data, and why Alexander's organic worldview constitutes the third view of space.

**1. Introduction**
Emerging geo-referenced big data from the Internet, particularly social media such as OpenStreetMap and Twitter, provides a new instrument for geospatial research. Big data shows some distinguishing features from small data (Mayer-Schonberger and Cukier 2013). For example, big data is accurately measured and individually based with geolocations and time stamps, rather than estimated and aggregated as small data. This makes big data unique and powerful for developing new insights into geographic forms and processes (e.g. Jiang and Miao 2015). On the other hand, big data poses enormous challenges in terms of data representation, structuring, and analytics. Unlike small data, big data is unstructured and massive, so conventional structured databases are unlikely to be of much use for data management. Conventional Gaussian statistics and Euclidean geometry are also not of much use for big data analytics (Jiang and Thill 2015, Jiang 2015b). Furthermore, conventional geographic representations, such as raster and vector, are inconvenient for providing deep insights into geographic forms and processes. For example, with raster and vector representations, geographic space is abstracted as either a large set of pixels or a variety of points, lines and polygons, layer by layer (Longley et al. 2015). Under the conventional geographic representations in general, geographic space is just a collection of numerous lifeless pieces, such as pixels, points, lines, and polygons, and there is little connection between these pieces, except for immediate, nearby relationships. These local relationships indicate that a geographic space is full of more or less similar things, and one can hardly see geographic space as a living structure, consisting of far more small things than large ones. Present geographic representations are very much influenced by the mechanistic worldview inherited from 300 years of science. Under the worldview, geographic space is mechanistically conceptualized as continuous fields or discrete objects (Couclelis 1992, Cova and Goodchild 2002), which pose little meaning in our minds and cognition. These representations, mainly based on Newton's absolute space and Leibniz's relational space, are essentially geometry based rather than topology oriented. By geometry, we mean geometric details such as locations, sizes, and directions, whereas the topology enables us to see the underlying living structure of far more small things than large ones. Thus the notion of topology differs fundamentally the same notion conceived and used in the geographic information systems (GIS) literature.

Geographic space as a whole is made of things – spatially coherent entities such as rivers, buildings, streets, and cities. Things are connected to other things, to constitute even larger things. For example, a set of streets or buildings constitutes a neighborhood, a set of neighborhoods constitutes a city, and a set of cities constitutes a country. These things have not yet become basic units of geographic representations in GIS (Longley et al. 2015). This is mainly due to the fact that current science is mainly mechanistic. The mechanistic worldview is remarkable and excellent, and all that we have achieved in science and technology is essentially based on this worldview. However, it is limited, in particular with respect to rebuilding architecture or making good built environments, as argued by Alexander (2002-2005). In order to build beautiful buildings, Alexander (2002-2005) conceived and developed a new worldview – a new conception of how the physical world is constituted. This new world picture is organic, so it differs fundamentally from the mechanistic one. Under the organic worldview, the world is an unbroken whole that possesses a physical structure, called a *living structure* or *wholeness* (see Section 2 for an introduction) Alexandrine organic space constitutes the third view of space, on which we will further discuss in Section 5.

Based on the third view of space, a topological representation has been previously developed (see Section 2 for an introduction) in order to show living structures in built environments and to further argue why the design principles of differentiation and adaptation are essential to reach living structures (Salingaros 2005, Jiang 2017). The present paper further explores how the topological representation or its illustrated living structure can be used to predict human activities. We will demonstrate, contrary to what we naively think, that geospatial big data is extremely well structured according to its underlying living structure, or the underlying scaling of far more small things than large ones. We will show that tweet locations can be well predicted by the living structure of street nodes extracted from OpenStreetMap. This predictability is not only based on the current status but also for the future status. We further illustrate why the topological representation is a truly multiscale representation, and subsequently put forward a new way of structuring geospatial data both big and small.

The remainder of this paper is structured as follows. Section 2 introduces the living structure or wholeness as a field of centers. Section 3 describes data and data processing. Section 4, based on the United Kingdom (UK) case studies, shows that tweet locations can be predicted using street nodes. Section 5 further discusses on the implication of the results, particularly how living structures of geographic space can be used to structure geospatial data. Finally, Section 6 draws conclusion and points to future work.

## 2. Living structure and the topological representation

Living structure is a key concept of this paper, developed by Alexander (2002-2005) in his Theory of Centers, and it is also called wholeness or life or beauty. The living structure consists of many individual centers that appear at the different levels of detail of the structure, and tend to overlap and nest within each other to form a coherent whole. The terms of living or life are not particularly in the biological sense, but in terms of the underlying recursive structure, – one with far more small things than large ones, one with numerous smallest things, a very few largest things, and some in between the smallest and largest. Alexander (2002-2005) identified 15 fundamental properties (Table 1) that help judge whether a thing is a living structure, or whether a thing is living or with life; usually the more properties the thing has, the more living the thing is. In this section, we first introduce the notion of living structure and its fundamental properties using an ornament as a working example (Figure 1), and then use a configuration of 10 fictive cities (Figure 2) to illustrate the topological representation and to show how the degree of livingness can be measured.

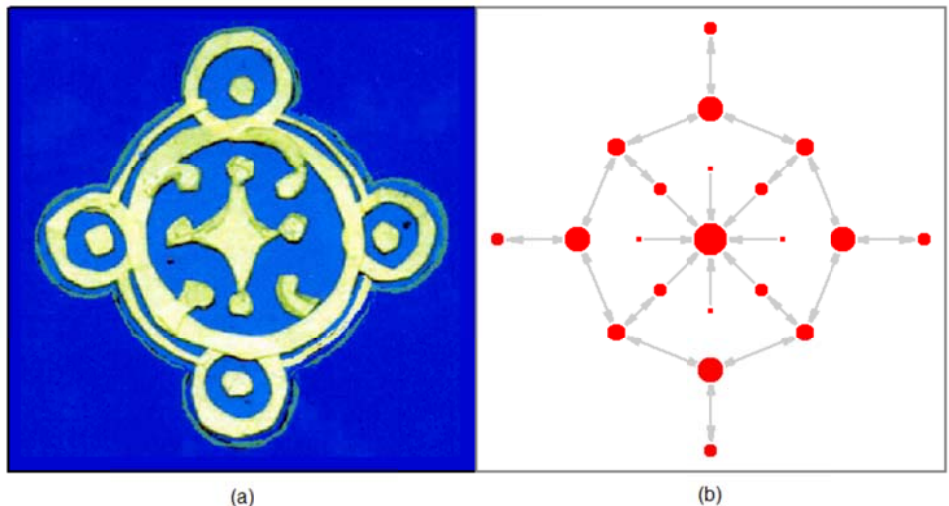

Figure 1: (Color online) An ornament and its topological representation
(Note: The ornament (a), presumably drawn by Alexander, appears in the book cover of *The Nature of Order* (Alexander 2002-2005). The dot sizes shown in the topological representation (b) illustrate the degree of livingness of the ornament structure: the larger the dots, the higher the degree of the livingness.)

The ornament is part of the book cover of *The Nature of Order* (Alexander 2002-2005), and it shows a strong sense of livingness, with far more small things than large ones (Figure 1a). The central theme of the four-volume book is about the nature of order, which is presumably represented by the big circle, while the four volumes, we suspect, are represented by four small circles. The four small circles are enhanced by the smaller dots within them, through so called differentiation processes. The big circle is further differentiated and therefore strengthened by the diamond-shaped piece, to which there are four dots attached. The big circle can be perceived as four arcs, which are further enhanced by the four little dots or strokes, as well as the four boundaries. The ornament possesses many of the 15 properties (Table 1). There are at least three levels of scale: big circle, small circles, and dots. There are many strong centers, and some are with thick boundaries. Alternating repetition is present, although less apparent, at the edge of figure and ground of the ornament. Good shapes repeat alternatively in the ornament, which is good shape itself, since it contains many good shapes in a recursive manner. Local symmetries are present with the diamond in the center, as well as in big circle with the four strokes. This ornament looks like a hand drafting, but we believe Alexander deliberately wanted to do so. He knew better than anyone else that roughness is such an important property of living structure. The diamond with four dots in the middle appears to echo the big circle with four small circles. All identified centers are not separate from each other, but tie together to become a coherent whole. This whole is topologically represented as a graph (Figure 1b).

Table 1: The 15 fundamental properties of the living structure (Alexander 2002-2005)

| | | |
|---|---|---|
| Levels of scale | Good shape | Roughness |
| Strong centers | Local symmetries | Echoes |
| Thick boundaries | Deep interlock and ambiguity | The void |
| Alternating repetition | Contrast | Simplicity and inner calm |
| Positive space | Gradients | Not separateness |

The topological representation, developed by Jiang (2017), is to build up supporting relationship among individual centers. For example, the big circle, consisting of four arcs, is supported by the four small circles, and the diamond is supported by the four attached dots. These supporting relationships are indicated by directed links in the topological representation of the ornament (Figure 1b). It should be noted that the topological representation does not show all potential centers. Interested readers can refer to Alexander (2002-2005) and Jiang (2016), in which a paper with a tiny dot induces up to 20 centers. With the topological representation or graph, and based on the mathematical model of wholeness (Jiang 2015a), we can compute degrees of livingness, as shown in Figure 1b. Geographic space is much more complex than the ornament, and its centers are much harder to identify than those of the ornament. Let's assume 10 fictive cities in a square space, and their sizes are respectively 1, 1/2, 1/3, … 1/10 (Figure 2a). The 10 cities can be put into 3 hierarchical levels based on the head/tail breaks classification (Jiang 2013). The three hierarchical levels are indicated by the three colors in Figure 2b, in which the dots of three different hierarchical levels are respectively used to create Thiessen polygons. A complex network is then created for all polygons, with directed relationships from smaller dots to adjacent larger ones at the same hierarchical level, and from contained polygons to containing ones between two consecutive levels (Figure 2c). With the network and the mathematical model of wholeness (Jiang 2015a), the degree of livingness can be obtained and visualized by the dot sizes in Figure 2c. In what follows, we will briefly introduce the ideas behind the mathematical model.

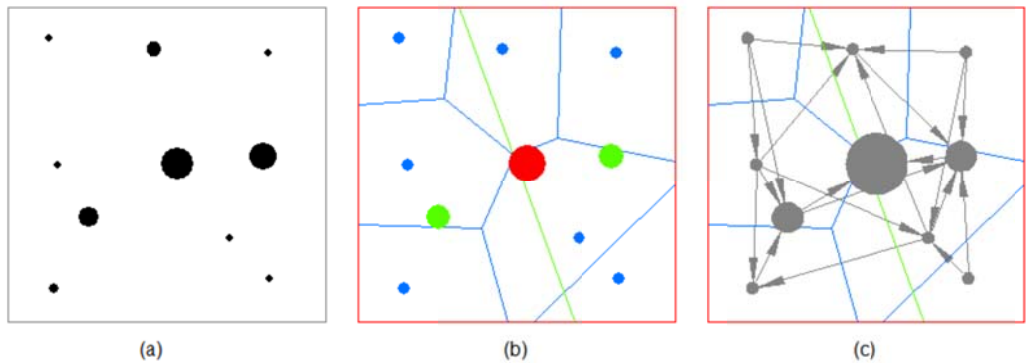

Figure 2: (Color online) Illustration of the topological representation
(Note: The 10 fictive cities with sizes of 1, 1/2, 1/3 ... and 1/10 are given some locations in a square space (a). These 10 cities are put into 3 hierarchical levels indicated by the 3 colors, and their corresponding Thiessen polygons nest each other (b). A complex network is then created (c) to capture adjacent relationships of the polygons at the same level, and nested relationships of the polygons across the levels.)

The set of the cities constitutes a living structure, and this living structure has two statuses: the current ($t_i$) and the future ($t_i+1$) shown in Figure 2b and 2c, respectively. In the current status, the degree of livingness is measured by city sizes, whereas in the future status, the degree of livingness is measured by Google's PageRank (PR) scores (Jiang 2015a). The major difference between these two measures lies in the configuration, i.e. how the 10 cities support each other to constitute a coherent whole. The configuration effect is continuous, which means that city sizes, as they are now, are outcome of the configuration, and the PR scores can be regarded as the future city sizes' ranking. For example, the two middle-sized cities are respectively supported by the three and the four small cities, and the largest city is supported by two middle-sized cities. For a spatial configuration that is well adapted, the two statuses or the two measures (sizes and PR scores) have little difference in terms of their individual ranking. Space, or spatial configuration to be specific, is not neutral, and it has the capacity to be more adapted or less adapted, or equivalently to be more whole or less whole. This dynamic view of space is what underlies Alexander's organic worldview that space is not lifeless or neutral, but a living structure involving far more small centers than large ones, and more importantly, space is in a continuous process of adaptation. This adaptation depends on not only how various centers adapt each other within their whole spatial configuration, but also beyond in terms of how it fits to its surroundings. For example, the degree of livingness of the ornament is not only decided by its centers within, but also influenced by its surroundings for a bigger whole of the book cover, and even beyond.

### 3. Data and data processing

In order to demonstrate that living structure can be used to predict human activities, or how human activities are shaped by the underlying living structure, we used two big datasets about the UK. They are street nodes and tweet locations between June 1–8, 2014 (Table 2). The street nodes were extracted from OpenStreetMap for building up natural cities (Jiang and Miao 2015) as a living structure, while tweet locations are used to verify if they can be predicted by the living structure. The street nodes refer to both street junctions and ending nodes. Street nodes can be easily derived if one writes a simple script to extract nodes with one street segment or at least three street segments in a street network. The nodes with one segment are ending points, while those with at least three street segments are junctions. For the convenience of the readers, we introduce how to use ArcGIS to derive street nodes from a network. There are two sets of procedures, which vary in efficiency and accuracy. The first is very accurate, but with low efficiency that is suitable for city-scale networks, while the second is less accurate, but with high efficiency that is suitable for country-scale networks.

The first set of procedures relies on ArcGIS' topology-building function, which partitions all streets on their junctions into different arcs. Under ArcToolbox Data Interoperability Tools, use Quick Export to create coverage format, which contains detailed topology. Use ArcToolbox Data Management Tool > Feature Vertices to Points to get both ends of these arcs, and then use Find Identical to get the number of ends at a same location. Those ends with the number of 1 or ≥ 3 are valid street nodes. The second set of procedures is based on ArcGIS' Intersect function, again within ArcToolbox. OSM street data must first be merged according to the same name. Then, use Intersect to get all junctions, and use Feature Vertices to Points to get the dangling end. Note that the Intersect operation can generate duplicate junctions that must be removed by Delete Identical to get all valid junctions. Finally, merge junctions and dangling ends to get all street nodes.

Table 2: The two datasets and derived natural cities or hotspots
(Note: OSM = Street nodes, Tweets = Tweet locations, NCities = Natural cities at different levels of scale, London II = The largest natural city within London, and London III = The largest natural city within London II)

|  | OSM | Tweets | NCities |
|---|---|---|---|
| UK | 4,715,279 | 2,933,153 | 123,551 |
| London | 308,999 | 424,970 | 16,080 |
| London II | 37,982 | 91,325 | 2,512 |
| London III | 1,929 | 2,796 | 89 |

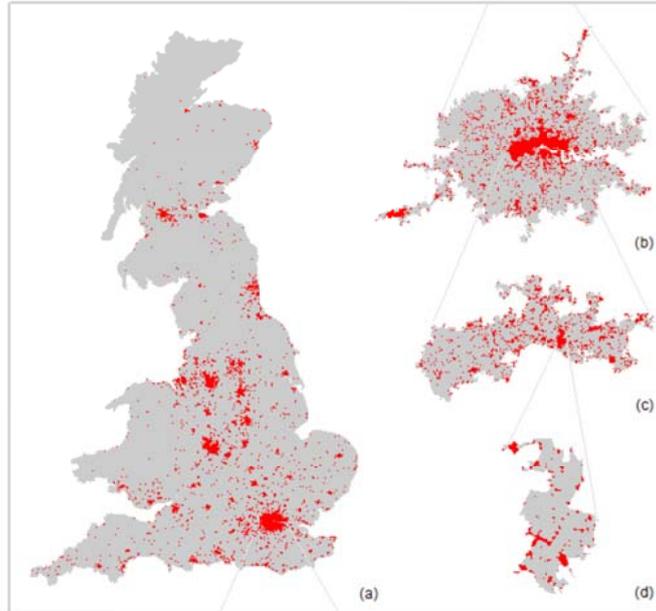

Figure 3: (Color online) Nested natural cities in the UK at different levels of scale
(Note: (a) UK and its natural cities, (b) London (a bit smaller than M25 highway) and its natural cities, (c) London II (very central at the northern side of Thames) and its natural cities, and (d) London III (near the city of London) and its natural cities, so a nested relationship as such: London III ⊂ London II ⊂ London ⊂ UK.)

The derived street nodes are used to generate natural cities at different levels of scale. Natural cities are naturally, objectively derived patches from big data, such as nighttime imagery, street nodes, points of interest, and tweet locations based on head/tail breaks (Jiang 2013). We first build a huge triangulated irregular network (TIN) by connecting all the street nodes or locations. This TIN comprises a large number of edges that are heavy-tail distributed, indicating far more short edges than long ones. Based on head/tail breaks, all the edges are put into two categories: those longer than the mean, called the *head*; and those shorter than the mean, called the *tail*. The edges in the tail, being shorter edges which imply higher density locations, eventually constitute individual patches called *natural cities* (see the Appendix in Jiang and Miao (2015) for a tutorial). It should be noted that the same notion of natural cities was used to refer to naturally evolved cities (Alexander 1965) rather than naturally derived cities used in this paper. However, the naturally derived cities are likely to resemble the naturally evolved cities, thus becoming an important instrument for studying city structure and dynamics; for example, the natural cities derived from the social media Brightkite check-in locations with time stamps (Jiang and Miao 2015) can be considered to be naturally evolved cities, in particular in terms of underlying mechanisms. Every natural city is a living center, and all natural cities of a country constitute an interconnected whole as a living structure (Jiang 2017). The next section will examine how tweet locations can be predicted by the living structure of geographic space at different levels of scale. Natural cities at the city scale could be called *hotspots*. However, for the sake of convenience, we will use *natural cities* only to refer to the naturally derived patches at different levels of scale. Figure 3 shows natural cities that are defined at the four levels of scale or spaces: London III, in London II, in London, and in the UK.

## 4. Prediction of tweet locations through living structure

The derived natural cities at the four spaces are topologically represented to set up complex networks, according to the principle illustrated in Figure 2. We built up four complex networks, respectively, for the four spaces, which are nested within each other (Figure 3). With these complex networks, we computed the degree of wholeness for both individual centers and their wholes. Seen in Figure 3, the four spaces are living structures, since each of them contain far more small natural cities than large ones. In what follows, we will examine how tweet locations can be well predicted by the living structures. Each of the living structures consists of far more small centers (represented as polygons) than large ones; see an example of London II (Figure 4a) in which both natural cities and their Thiessen polygons act as centers (Figure 4b for the enlarged view). The centers we refer to here consist of both Thiessen polygons and natural cities, while the previous study by Jiang (2017) referred to only Thiessen polygons. The natural cities act as the cores of their corresponding Thiessen polygons, reflecting the centeredness of the centers (Alexander 2002-2005).

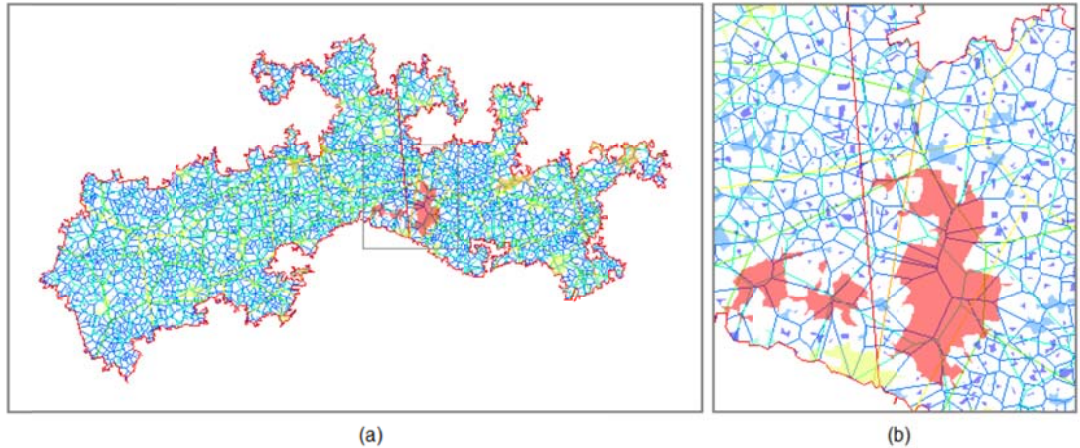

Figure 4: (Color online) The topological representation of London II
(Note: London II demonstrates a living structure with a very striking scaling hierarchy of far more small polygons than large ones: (a) an overall view of London II, and (b) an enlarged view around the largest natural city of London II. The spectral color legend is used to indicate the living structure of far more small polygons (cold colors) than large ones (warm colors), or numerous smallest (blue), a very few largest (red), and some in between the smallest and largest (other colors between blue and red).)

### 4.1 Correlations at the scale of Thiessen polygons
We found good correlations between living structure and tweet locations, with R square values ranging from 0.63 to 0.99 (Table 3). The column "OSM/Tweets" indicates the current status, while the column "Life/Tweets" indicates the future status. The good correlations imply that living structures can predict tweet locations at the scale of Thiessen polygons. Figure 5 roughly illustrates how well street nodes correlate tweet locations. However, we are not used to this view of space or this kind of multiscale spatial units that are nested within each other. This is somehow like a series of spatial units involving country, states, counties, with small units contained in the big ones. We are used to single-scale units, either all states or all countries, rather than mixed units of states and counties. It is exactly the multiscale units that make the topological representation unique and powerful, since it captures the scaling or living structure of space. Besides the correlation at the nested polygons scale, we could also examine the prediction at the scale of the natural cities.

Table 3: R square values among street nodes, tweet locations, and degree of wholeness at the scale of Thiessen polygons
(Note: OSM = Street nodes, Tweets = Tweet locations, Life = Degrees of wholeness, and / = Between)

|            | OSM/Tweets | Life/Tweets |
|------------|------------|-------------|
| UK         | 0.99       | 0.76        |
| London     | 0.98       | 0.81        |
| London II  | 0.92       | 0.63        |
| London III | 0.97       | 0.85        |

### 4.2 Correlations at the scale of natural cities
Good correlations also occur at the scale of natural cities. Table 4 presents the results of R square of different pairs and at different levels, although not as good as those at the level of polygons. It should be noted that both the percentages of street nodes and tweet locations included in natural cities dramatically decrease. For example, the percentage of tweets decreases from 84 percent at the country scale to 18 percent at the London III scale. This is because tweets are more evenly distributed in the natural cities than in the country. Alternatively, tweets are more heterogeneously distributed in the country than in the natural cities. The living structure or the kind of topological analysis can well predict tweet locations, but only for those within natural cities. The problem is that a vast majority of tweets are not within natural cities. In this case, we do not recommend using natural cities, but instead Thiessen polygons for effective prediction. In other words, natural cities can predict those tweets that are highly clustered or concentrated, while Thiessen polygons can be used for all tweets both highly clustered and highly segregated.

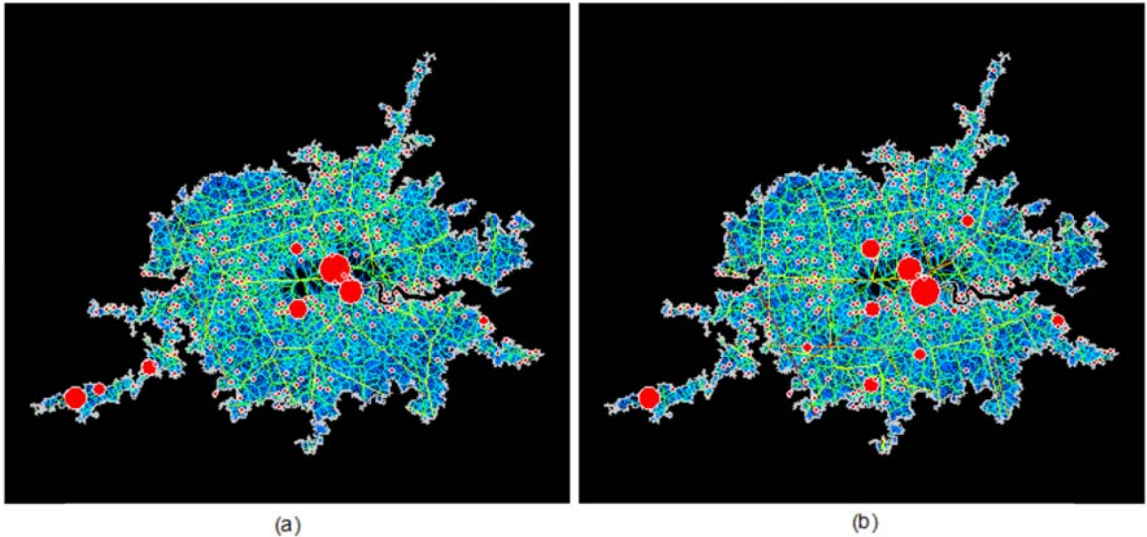

Figure 5: (Color online) Illustration of good correlation between the polygon sizes, based on street nodes (a) and their degrees of wholeness (b) for the natural city London (a bit smaller than M25 highway)
(Note: The sizes of the dots are proportional to their values, so they are not classified in order to show the good correlation.)

Table 4: Percentages of data within natural cities and R square values among street nodes, tweet locations, and degree of wholeness at the scale of natural cities
(Note: OSM% = Street nodes included in natural cities, Tweets% = Tweet locations included in natural cities, OSM = street nodes, Tweets = Tweet locations, Life = degree of wholeness, and / = between)

|  | OSM% | Tweets% | OSM/Tweets | Life/Tweets |
|---|---|---|---|---|
| UK | 0.89 | 0.84 | 0.93 | 0.55 |
| London | 0.84 | 0.40 | 0.85 | 0.44 |
| London II | 0.86 | 0.24 | 0.70 | 0.58 |
| London III | 0.55 | 0.18 | 0.84 | 0.64 |

## 4.3 Degrees of wholeness or life or beauty

The UK as a whole, and its sub-wholes, such as London, London II and London III, are living structures. It would be interesting to compare their degrees of wholeness, life, or beauty. For this purpose, we computed correlations between street nodes and wholeness of the natural cities (Table 5; see the columns OSM/Life (TS) and OSM/Life (NC) and their averages, shown in column OSM/Life). The UK has the highest degrees of wholeness, indicating that the country is more beautiful than its cities. On the other hand, the UK, London, London II and London III all have the same degree of adaptation, given the same correlation between cities sizes and their degrees of wholeness.

Table 5: Comparison of wholeness among the UK, London, London II and London III
(Note: OSM = street nodes, Tweets = Tweet locations, Life = degree of wholeness, / = between, TS = Thiessen polygons, and NC = Natural cities)

|            | Ht-index | OSM/Life (TS) | OSM/Life (NC) | OSM/Life |
|------------|----------|---------------|---------------|----------|
| UK         | 8        | 0.78          | 0.75          | 0.77     |
| London     | 6        | 0.85          | 0.65          | 0.75     |
| London II  | 6        | 0.76          | 0.79          | 0.77     |
| London III | 4        | 0.86          | 0.63          | 0.75     |

It should be noted that all the correlations examined above are significant at the 0.01 level (two-tailed) based on the logarithmic scale of the data. Through these computing results, we can understand why the UK is a living structure because its constituents are living. To be more specific, the country is living because its constituents (e.g. London) are living; London is living, because its centers (e.g. London II) are living. The reason that London II is living is because its centers (e.g. London III) are living. More generally, goodness of things is based on the recursive way of assessment. This is what underlies the notion of living structures. From this case study, we have seen already how living structure is extracted from big data of street nodes, and how subsequently human activities are shaped by the living structure. Big data differs from small data, so big data is better than small data in capturing the underlying configuration of geographic space.

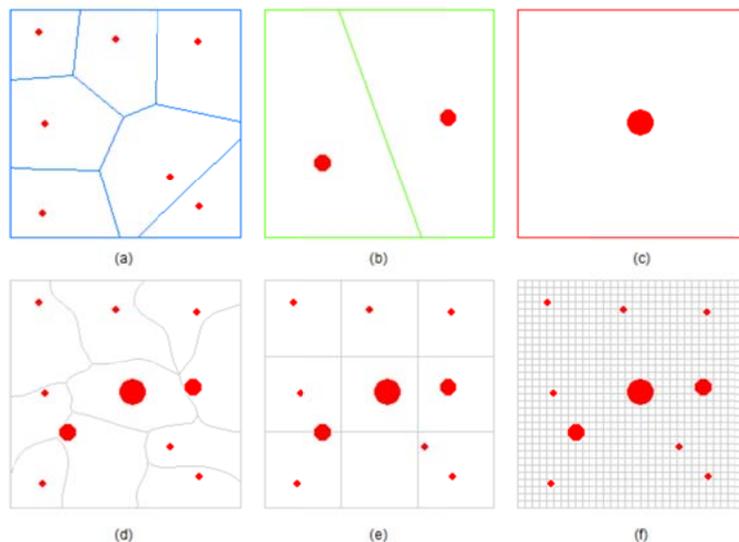

Figure 6: (Color online) The topological representation as a multiscale representation versus single-scale representations commonly used in GIS
(Note: To illustrate, the topological representation is partitioned into three scales, ranging from (a) the smallest, to (b) the medium, to (c) the largest. There are three single-scale representations: (d) administrative boundaries, (e) a regular grid, and (f) an image or a pixel-based representation.

## 5. Implications on the topological representation and living structure

Through the above case studies, we have seen that space is not lifeless or neutral, but a living structure involving far more small centers than large ones. In this regard, the topological representation is shown to be efficient and effective for illustrating the underlying living or scaling structure. The topological representation is a multiscale representation – multiple scales ranging from the smallest scale to the largest. To illustrate, we partition the representation in Figure 2 into three scales in Figure 6 (Panels a, b, and c). However, existing geographic representations such as raster and vector are essentially single scale, reflecting mechanistic views of space of Newton and Leibniz (see Panels d, e, and f of Figure 6). Single-scale representations create many scale-related problems and have been major concerns in geographical analysis, such as the modifiable areal unit problem (Gehlke and Biehl 1934, Openshaw 1984), the conundrum of length (Richardson 1961, Mandelbrot 1982, Batty and Longley 1994, Frankhauser 1994, Chen 2008), and the ecological fallacy (Robinson 1950, King 1997, Wu et al. 2006). The single-scale representations are suitable for showing geographic features of more or less similar scales. Therefore, it is not surprising that current spatial statistics focuses much on autocorrelation, little on scaling or fractal or living property of far more small geographic features than large ones (Jiang 2015b). The topological representation enables us to see not only more or less similar things in one scale (spatial dependency), but also far more small things than large ones across all scales (spatial heterogeneity).

The notion of far more small things than large ones or spatial heterogeneity has been formulated as a scaling law (Jiang 2015b). It is complementary to Tobler's law (1970) or the first law of geography for characterizing geographic space or the Earth's surface: scaling law being global, while Tobler's law being local. The scaling law only refers to statistical property, without referring to the underlying geometrical property. This is essentially the same as Zipf's Law (1949) about city-size distribution, without referring to spatial configuration of cities formulated by the Central Place Theory (Christaller 1933). However, the Theory of Centers (Alexander 2002–2005) about living structure concerns both the statistical and geometrical aspects. The statistical aspect illustrates the fact that city sizes meet a power-law relationship, while the geometrical aspect points to the fact that all cities adapt to each other to form a coherent whole (Jiang 2017). Seen from both statistical and geometric aspects, space is far more heterogeneous than what current spatial statistics and Euclidean geometry can effectively deal with. It is in this sense that we must adopt fractal geometry and Paretian statistics for geospatial analysis. It is in this sense that we must adopt the topological representation for getting insights into the living structure of space. It is in this sense that existing geographic representations show critical limitations.

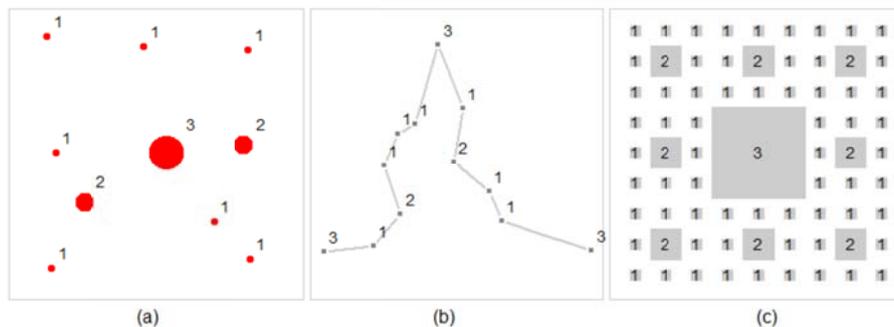

Figure 7: (Color online) Structuring geographic features based on their scaling hierarchy
(Note: Geographic features of different shapes: (a) points, (b) lines, and (c) polygons, in which the ht-indices or numbers indicate their hierarchical levels.)

Given the limitations of single-scale representations, topological representation (or the underlying living structure) is a better alternative for representing or structuring geospatial data. The hierarchical data structures, such as quadtree (Samet 2006), should be adopted to reflect the living structure of geographic space or features. All geographic features should be put into a whole, and their status can be indicated by their hierarchical levels, or their ht-index within the whole. Figure 7 illustrates three kinds of geographic features, which all demonstrate living structures. A set of far more small cities than large ones constitute a coherent whole (Figure 7a), and a curve is regarded as a set of far more small bends than large ones (Figure 7b), rather than a set of more or less similar segments. The Sierpinski carpet (Figure 7c) is a proxy of many areal geographic features, such as islands, lakes, and land use patches. A geographic space or feature is well structured hierarchically, so mechanistically imposed representations, such as raster and vector, are not appropriate for illustrating its living structure. Therefore, administrative boundaries, regular grids, or pixel-based images are not appropriate for revealing the living structure of geographic features. In this regard, naturally or organically derived cities provide a significant instrument for structuring big data.

To this point, we can further elaborate on how Alexander's organic worldview constitutes the third view of space. In the history of science, there are two dominant conceptions of space that are also called absolute and relative views of space. The absolute view arises out of Newtonian physics, and implies that phenomena can be defined in themselves, so a space can be considered to be a container. The relative view of space comes from Leibniz's conception of space that space can be defined as the set of all possible relationships among phenomena. These two views of space reflect pretty well the conception of space by Descartes (1954), who described space as a neutral and strictly abstract geometric medium, through his uniform spatial scheme of analytical or coordinate geometry. This spatial scheme, similar to the common geographic representations raster and vector, made us to think of space as a neutral, lifeless, and dead substance. Alexander (2002-2005) challenged this mechanistic conception of space, and conceived an organic worldview, under which space has capacity to be more living or less living according to its inherent structure. Unlike absolute or relative space, this new view of space is organic, based on the concept of wholeness, which finds its roots in many disciplines such as quantum physics (Bohm 1980), and Gestalt psychology (Köhler 1947). Alexander (2002-2005) further argues the ubiquity of wholeness in nature, in buildings, in works of art, and more specifically in any part of space at different levels of scale. Eventually, the goodness of a given part of space may be understood only as a consequence of the wholeness that exists. In other words, the wholeness is the essence of geographic space.

Geography as a science has three fundamental issues to address about geographic space: (1) how it looks, (2) how it works, and (3) what it ought to be. The first issue concerns mainly about geographic forms or urban structure, which are governed by two fundamental laws: scaling law (Jiang 2015b) and Tobler's Law (1970). Scaling law states that there are far more small things than large ones on the Earth's surface, whereas Tobler's Law refers to the fact that similar things tend to be nearby or related. Kriging interpolation is possible because of Tobler's law. Prediction of tweet locations is possible because of scaling law or living structure. The second issue refers to the underlying mechanisms in terms of how a complex or living structure evolves. This is what physicists are primarily concerned with; for example, how cities evolve, and how topographic surfaces are formed with respect to geological processes. The surface complexity arises out of the deep simplicity – the deep nonlinear and chaotic processes (Gribbin 2004). In other words, geographic processes or urban dynamics in particular fluctuate very much like stock prices. The first two issues are fundamental to many other sciences, such as physics, biology, and chemistry, for understanding and explaining complex natural and societal phenomena. The third issue is not so common to other sciences, which are hardly concerned with creation or design (Alexander 2002-2005). The issue of what it ought to be intends to address how to create a living structure, and how to make a living structure more living or more harmonic. This third issue has not been well addressed in geography, yet it is so unique and important in terms of how to make better or more sustainable built environments. As demonstrated by Alexander (2002-2005), the concept of living structure has paved a way to this creation and design of living environments, through harmony-seeking computation (Alexander 2009). In this connection, harmony-seeking computation as a kind of adaptive computation deserves further research in the future.

## 6. Conclusion

Inspired by Alexander's new cosmology, this paper demonstrated that human activities, such as tweet locations, can be well predicted by the underlying living structure using topological representation and analysis. From this study, we have a better sense of understanding how human activities are shaped by space, or more precisely by its underlying living structure. This finding further validates topological representation as an effective tool for geospatial analysis, particularly in the context of big data. More importantly, we showed that living structure exists in space, to varying degrees at different levels of scale, so it is a legitimate object of inquiry for better understanding goodness of built environments. Unlike existing geographic representations, such as raster and vector, which are essentially single-scale, the topological representation is a de-facto multiscale representation – multiple scales, ranging from the smallest to the largest and including some in between, within the single representation. This multiscale representation can help avoid many scale issues caused by the traditional single-scale representations.

Big data provides not only a new type of data sources for geographic research, but also poses a big challenge on how to efficiently and effectively manage it, particularly how to develop new penetrating insights. Unlike small data, which rely on samples for understanding the population of the data, big data is able to effectively extract living structure of space. Big data is more representative than small data in capturing the underlying living structure. This is why big data or living structure can effectively predict human activities. From another aspect, big data is commonly considered to be unstructured, yet the underlying living structure makes big data incredibly well structured. This is because space itself is inherently well structured. It is in this context we suggest a new way of structuring big data through living structure. The living structure, and in particular its current and future statuses, indicate that space is living rather than lifeless, and space is dynamics rather than static. The essence of geographic space is its living structure. The living structure can be used to structure geospatial data by relying on its inherent hierarchy. This proposal deserves further research in the future.

**Acknowledgements**
We would like to thank the three anonymous reviewers for their constructive comments. We also would like to thank Junjun Yin for kindly sharing the tweet location data.